\documentclass[preprint]{aastex}
\usepackage{hyperref}
\shorttitle{Spitzer Spectroscopy of 2MASS J04414489+2301513}
\shortauthors{Adame et al.} 

\begin{document}

\title{\emph{Spitzer} Spectroscopy of the Circumprimary Disk in the
Binary Brown Dwarf 2MASS J04414489+2301513\altaffilmark{1}}

\author{
Luc\'ia Adame\altaffilmark{2,3},
Nuria Calvet\altaffilmark{3},
K. L. Luhman\altaffilmark{4,5},
Paola D'Alessio\altaffilmark{6},
Elise Furlan\altaffilmark{7},
M. K. McClure\altaffilmark{3},
Lee Hartmann\altaffilmark{3},
William J. Forrest\altaffilmark{8},
and Dan M. Watson\altaffilmark{8}
}

\altaffiltext{1}{Based on observations made with the Spitzer Space
Telescope, which is operated by the Jet Propulsion Laboratory,
California Institute of Technology under a contract with NASA.} 

\altaffiltext{2}{Instituto de Astronom\'ia, Universidad Nacional
Aut\'onoma de M\'exico, AP 70-264, 04510 M\'exico DF, M\'exico.}

\altaffiltext{3}{Current address: Department of Astronomy, University of
Michigan, Ann Arbor, MI 48109; adamel@umich.edu}

\altaffiltext{4}{Department of Astronomy and Astrophysics, The
Pennsylvania State University, University Park, PA 16802.}

\altaffiltext{5}{Center for Exoplanets and Habitable Worlds, The
Pennsylvania State University, University Park, PA 16802.}

\altaffiltext{6}{Centro de Radioastronom\'ia y Astrof\'isica,
Universidad Nacional Aut\'onoma de M\'exico, Apartado Postal 72-3
(Xangari), 58089, Morelia, Michoac\'an, M\'exico.}

\altaffiltext{7}{JPL, Caltech, Mail Stop 264-767, 4800 Oak Grove Dr.,
Pasadena, CA 91109, USA}

\altaffiltext{8}{Department of Physics and Astronomy, The University of
Rochester, Rochester, NY 14627.}

\begin{abstract} 

Using the {\it Spitzer} Infrared Spectrograph, we have performed
mid-infrared spectroscopy on the young binary brown dwarf
2MASS~J04414489+2301513 (15 AU) in the Taurus star-forming region. The
spectrum exhibits excess continuum emission that likely arises from a
circumstellar disk around the primary. Silicate emission is not detected
in these data, indicating the presence of significant grain growth. This
is one of the few brown dwarf disks at such a young age ($\sim1$~Myr)
that has been found to lack silicate emission. To quantitatively
constrain the properties of the disk, we have compared the spectral
energy distribution of 2MASS~J04414489+2301513 to the predictions of our
vertical structure codes for irradiated accretion disks. Our models
suggest that the remaining atmospheric grains of moderately depleted
layers may have grown to a size of $\gtrsim5$~\micron. In addition, our
model fits indicate an outer radius of 0.2--0.3~AU for the disk. The
small size of this circumprimary disk could be due to truncation by the
secondary. The absence of an outer disk containing a reservoir of small,
primordial grains, combined with a weak turbulent mechanism, may be
responsible for the advanced grain growth in this disk.

\end{abstract} 


\maketitle

\section{Introduction}

Grain growth, crystallization, and settling occur in primordial
circumstellar disks around both stellar and substellar objects. These
processes produce several observational signatures. The mid-infrared
(IR) silicate emission becomes weaker as the grains in the upper layers
grow and settle towards the midplane while crystallization modifies the
shape of the band. IR continuum emission is reduced at longer
wavelengths with grain growth and settling because the intercepted
stellar flux decreases as the disk flattens. In the substellar domain,
these hallmarks of dust evolution have been reported previously for
disks at ages $\tau\sim1-10$~Myr
\citep{Apai04,Apai05,Scholz07,Morrow08,Pascucci09,Riaz09,RLG09}. For
instance, disks around the youngest brown dwarfs exhibit evidence of
small, crystalline silicate grains \citep[][]{Apai05,Riaz09} while older
brown dwarf disks have weaker silicate emission that indicates the
presence of large grains \citep{Scholz07,Morrow08,RLG09}. The evolution
of grains in the inner annuli of these disks, e.g., the zone where the
10~\micron\ silicate band emerges \citep[within
$10^{-3}-0.1$~AU,][]{KS07}, appears to occur more rapidly than in disks
around solar-mass stars. One possible explanation for this phenomena is
that the low levels of turbulence for brown dwarf disks produce lower
mass accretion rates and reduce the replenishment of the inner disk with
primordial grains from outer annuli \citep{SA07}. Also, the collisional
rate between large size grains would be small, preventing the
fragmentation to small size grains.

During a mid-IR spectroscopic survey of young brown dwarfs with the {\it
Spitzer Space Telescope} \citep{wer04}, we have identified a brown dwarf
disk that shows evidence of advanced grain evolution at a relatively
young age of $\sim1$~Myr. This brown dwarf, \object{2MASS
J04414489+2301513} \citep[][henceforth 2M~J04414489]{Luhman06}, lies in
the Taurus star-forming region and was recently shown to harbor a
companion with a projected separation of $0\farcs105$ ($\sim15$~AU) and
a mass of 5--10~$M_{\rm Jup}$ \citep{TLM10}. In this Letter, we use our
mid-IR spectrum of this system and our models of settled, irradiated
accretion disks to constrain the properties of the circumstellar disk
that resides around the primary.

\section{Observations}\label{obs}

We observed 2M~J04414489 with the {\it Spitzer} Infrared Spectrograph
\citep[IRS;][]{H04} on 2009 April 20 as a part of the Guaranteed Time
Observations of the IRS instrument team. We collected data with both of
the low-resolution IRS modules, Short-Low and Long-Low, which cover
5.3-14 and 14-40~\micron , respectively, with a resolution of
$\lambda/\Delta \lambda \sim90$. We used the basic calibrated data
produced by version S18.7 of the Spitzer Science Center pipeline. Before
extraction of the spectrum, bad and ``rogue" pixels were fixed by
interpolating adjacent, good pixels in the dispersion direction.
2M~J04414489 was then extracted using the optimal extraction tool in
SMART \citep{Higdon04,Lebouteiller10}. We chose the default calibration,
which is based on a relative spectral response function derived from
three calibrator stars. The young low-mass star 2MASS J04414565+2301580
is $12\farcs4$ from 2M~J04414489 and fell within the slit during the LL
observations. We performed optimal extraction at the nominal nod
position of 2M~J04414489 to extract its spectrum, which should result in
minimal contamination from 2MASS J04414565+2301580. Indeed, the
resulting flux levels of the data at SL and LL match at 14~\micron\
without applying any scaling factors. The IRS spectrum includes the flux
of both 2M J04414565 A and B, but it should be dominated by the former,
as explained in Section \ref{s31}.

\section{Analysis}

\subsection{Disk Emission}\label{s31}

\begin{figure}
\epsscale{1}
\plotone{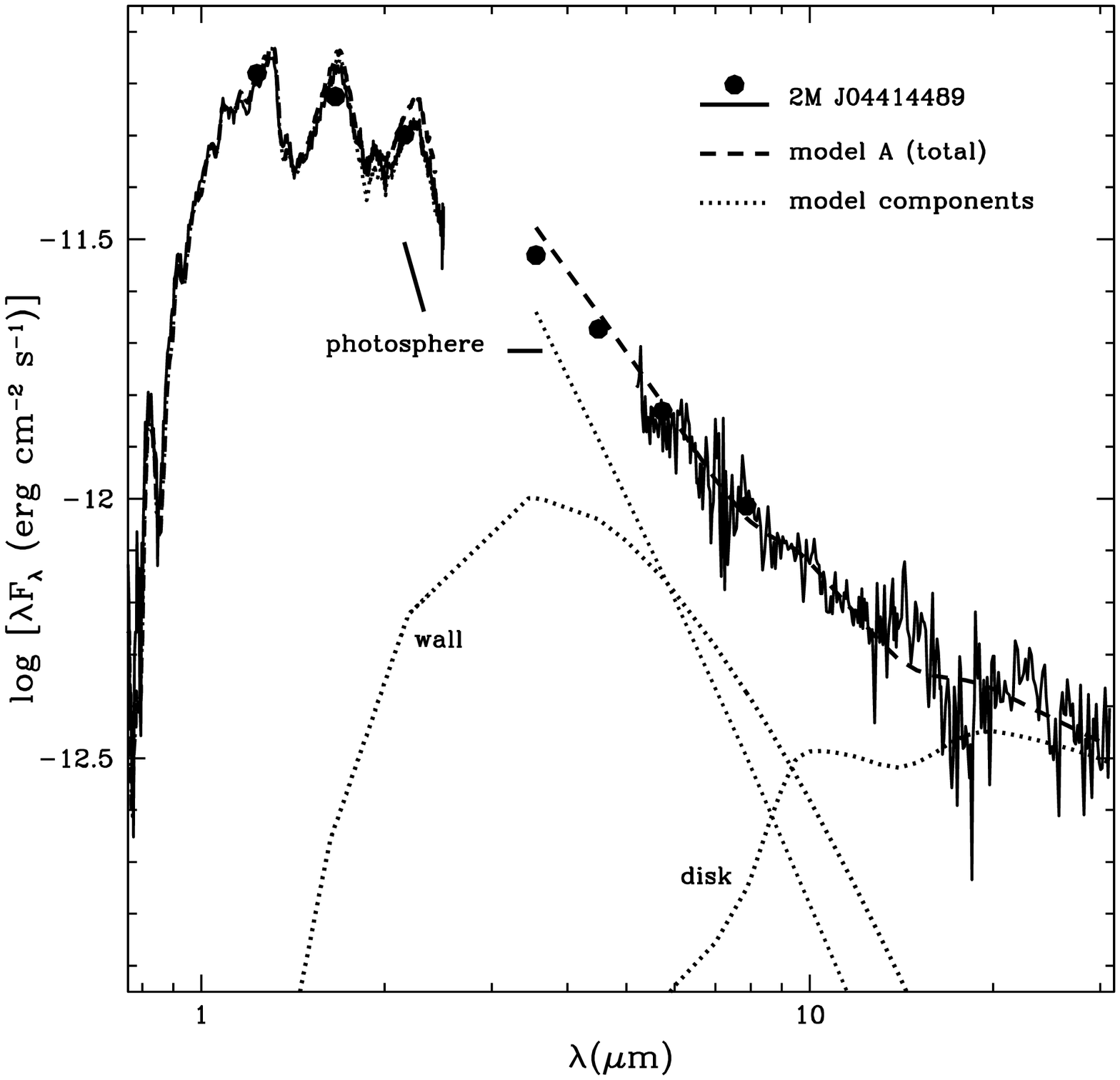}
\caption{SED of the young brown dwarf 2M~J04414489 compared to a model
of its circumstellar disk (dotted and dashed lines, model A from
Table~\ref{tbl-1}). We have adopted the SED of 2MASSW~J1139511-315921
\citep{Morrow08} to represent the stellar photosphere and have scaled
it to the 1-2~\micron\ flux of 2M~J04414489. For
$\lambda\le2.5$~\micron, the stellar photosphere overlaps almost
precisely with the photosphere of 2M~J04414489.}
\label{fig:fig1}

\end{figure}

We show the spectral energy distribution (SED) of 2M~J04414489 in
Figure~\ref{fig:fig1}, which was constructed with our IRS data,
0.8--2.5~\micron\ spectroscopy from \citet{Luhman06}, and 3--8~\micron\
photometry from \citet{Luhman10}. We have included in
Figure~\ref{fig:fig1} the SED for a diskless young brown dwarf near the
same spectral type as 2M~J04414489
\citep[2MASSW~J1139511-315921,][]{Morrow08}. Relative to the stellar
photosphere of the latter object, the SED of 2M~J04414489 exhibits
significant excess emission that indicates the presence of a dusty,
optically thick disk, which is consistent with earlier results based on
the mid-IR photometry \citep{Luhman10}. The disk is probably located
around the primary because a disk around the secondary would be too
faint to account for the observed excess emission based on the flux
ratio at near-IR wavelengths \citep[$\Delta K_s=1.54$,][]{TLM10} and the
typical near- to mid-IR colors of brown dwarfs with disks
\citep{Luhman10}. A circumbinary disk is not plausible since its inner
hole would be too large for the disk to produce significant mid-IR
emission. Finally, the mid-IR excess emission cannot be due to the
photosphere of the secondary since its IR colors should be similar to
those of the primary's photosphere. For instance, given the spectral
type of M8.5 for the primary and an expected type of M9.5-L0 for the
secondary \citep{Luhman06,TLM10}, their photospheres should have colors
near $K_s-[4.5]=0.71$ and 0.83, respectively \citep{Luhman10}. Combining
those colors with the flux ratio at $K_s$ suggests that the secondary's
photosphere produces an excess of only $\sim$0.02~mag in $K_s-[4.5]$ for
the system relative to the photospheric colors of the primary.

\subsection{Model Parameters}

The optical spectrum of 2M~J04414489 exhibits H$\alpha$ emission
\citep{Luhman06}, indicating the presence of ongoing mass accretion.
Therefore, we have modeled the mid-IR excess emission from 2M~J04414489
in terms of emission from a settled, irradiated accretion disk following
the procedures of \citet{Paola98, Paola05,Paola06} adapted for
substellar objects \citep{tesis}. For the stellar photosphere, we adopt
an effective temperature of 2555~K, which corresponds to its M8.5
spectral type \citep{Luhman03,Luhman06}. We also adopt a luminosity of
0.004~$L_{\sun}$ based on its $J$ magnitude and a mass of
0.025~M$_{\odot}$ based on theoretical evolutionary models
\citep{CBAH00}. In our models, we solve the vertical structure of the
disk while assuming a constant turbulent-viscosity parameter $\alpha$
and a uniform mass accretion rate, $\dot{M}$, through the disk. For most
of the brown dwarfs in Taurus, assuming $\alpha\ge10^{-3}$ is suitable
\citep{tesis}, although we allow $\alpha$ to range from $10^{-5}$ to
0.01 for our models of 2M~J04414489. In each annulus, the mass surface
density depend on the ratio $\dot{M}/\alpha$. We compute models with
$\dot{M}=10^{-12}$, $10^{-11}$, and $10^{-10}$~M$_{\sun}$~yr$^{-1}$,
which are typical values measured for young substellar objects
\citep{Natta04,Muzerolle05,MJB05,HH08,Herczeg09}.

The adopted dust model is composed of segregated spheres of astronomical
silicates and graphite with abundances and optical constants from
\citet{DL84} and \citet{WD01} and a size distribution of $n(a)\sim
a^{-p}$ \citep{MRN77}, where $p=3.5$. To account for grain growth, we
have varied the maximum grain size in the upper layers and the inner
wall, adopting values of $a_{max}=0.25$, 1, 5, 10, 100 and 1000~\micron
. Meanwhile, we assume $a_{max}=1$~mm for the dust population near the
midplane. As grains grow and settle towards the midplane, the atmosphere
is depleted of small grains. The model of \citet{Paola06} quantifies
this depletion through the parameter $\epsilon$, which is the
dust-to-gas mass ratio of the atmosphere relative to the standard ratio
\citep[with $\zeta_{sil}=0.004$ and $\zeta_{graph}=0.0025$,][]{DL84}. We
have considered $\epsilon=0.001$, 0.01, and 0.1 in our models. 

We also calculate the structure and SED of the inner wall , which is
heated by radiation from the stellar photosphere \citep{Paola05}. The
wall is located at radius $R_{in}$, which can be the dust destruction
radius (assuming a sublimation temperature of $\sim1400$~K) or the
magnetospheric radius, whichever is larger. We explore maximum dust
temperatures of $\sim1400$~K to 1000~K (in steps of 100~K),
corresponding to an inner dust gap with $R_{in}\sim3$ to 6~R$_{*}$. The
outer disk radius $R_d$ may be set by interaction with the secondary
\citep{PP77,AL94}. Assuming a mass ratio of the secondary to the primary
of $\sim0.3$, a projected semimajor axis of $\sim15$~AU \citep{TLM10},
and a mean Reynolds number of $10^{4}-10^{5}$ \citep[see][]{AL94}, the
truncation radius is predicted to be $3\lesssim
R_{truncation}\lesssim7$~AU for eccentricities between $e=0$ and
$e=0.5-0.6$ (a large circumprimary radius corresponds to a small
eccentricity). Therefore, we fixed $R_d$ to an initial value of 5~AU;
for promising models, we have selected values of $R_d$ that range from
0.1--5~AU. Finally, we adopt inclination angles of $i=10$, 20, and
$60^{\circ}$. The disk SED library consists of some 700 models.

\subsection{Model Fits}

We proceed to identify the combinations of disk parameters that can
reproduce the SED of 2M~J04414489. We perform a reduced-$\chi^{2}$
goodness-of-fit test between each model and the IRS data
\citep[e.g.][]{vanBoekel05}, where $\chi_r^{2}\sim1$ indicates a good
match. We find that models with $i=20^{\circ}$,
$10^{-11}\le\dot{M}\le10^{-10}$~M$_{\sun}$ yr$^{-1}$, and a wall
temperature of $\sim1000$~K match the SED at $\lambda<10$~\micron\
better than other values of these parameters. The radius of this inner
wall is $R_{in}\approx6$~R$_{*}\approx0.009$~AU. We allow the wall
height $z_{wall}$ to vary from $3.4$~H$_p$ to $4.1$~H$_p$, where H$_p$
is the gas scale height at $R_{in}$. The models with these values of
$i$, $\dot{M}$, and $R_{in}$ that produce $\chi_r^2<4$ are 21. An
important constrain for our model SED selection is the value of the mass
accretion rate. From the $\dot{M}-M_*$ relationship of \citet{HH08}, the
mass accretion rate onto this low mass brown dwarf would be
$\log{\dot{M}}\approx-10.7\pm0.6$, thus models with
$10^{-10}$~M$_{\sun}$ yr$^{-1}$ are in the higher end of the expected
mass accretion rate. On the other hand, the equivalent width of the
H$\alpha$ emission line of 2M~J04414489 \citep{Luhman06} is 100~\AA,
which corresponds to a 10\% width of H$\alpha$ of $\sim220$ km s$^{-1}$
\citep{Muzerolle05}, and to a mass accretion rate of
$\lesssim10^{-11}$~M$_{\sun}$ yr$^{-1}$ \citep{Natta04}. Hence we
discard the high mass accretion models, and select the remaining four
models as our best-fit models, which are listed in Table~\ref{tbl-1}.

The SED predicted by model A, which has the lowest value of $\chi_r^2$,
is compared to the observed SED in Figure~\ref{fig:fig1}. The variations
in the disk parameters among the best-fit models in Table~\ref{tbl-1}
are qualitatively explained by the fact that the mid-IR emission is
weaker for smaller $\epsilon$ and $\dot{M}$ and larger $a_{max}$ and
$\alpha$, causing a reduced optical depth. For instance, the grain
growth of model C and the depletion of model D must be balanced by
increasing the dust mass content in order for the models to produce
sufficient emission to match the data. Meanwhile, models with
$a_{max}=1$~mm and most of the models with
$10^{-12}\le\dot{M}\le10^{-11}$~M$_{\sun}$ yr$^{-1}$ produced emission
that was too weak to reproduce the data ($\chi^2_r\gg4$).

\begin{figure}
\epsscale{1}
\plotone{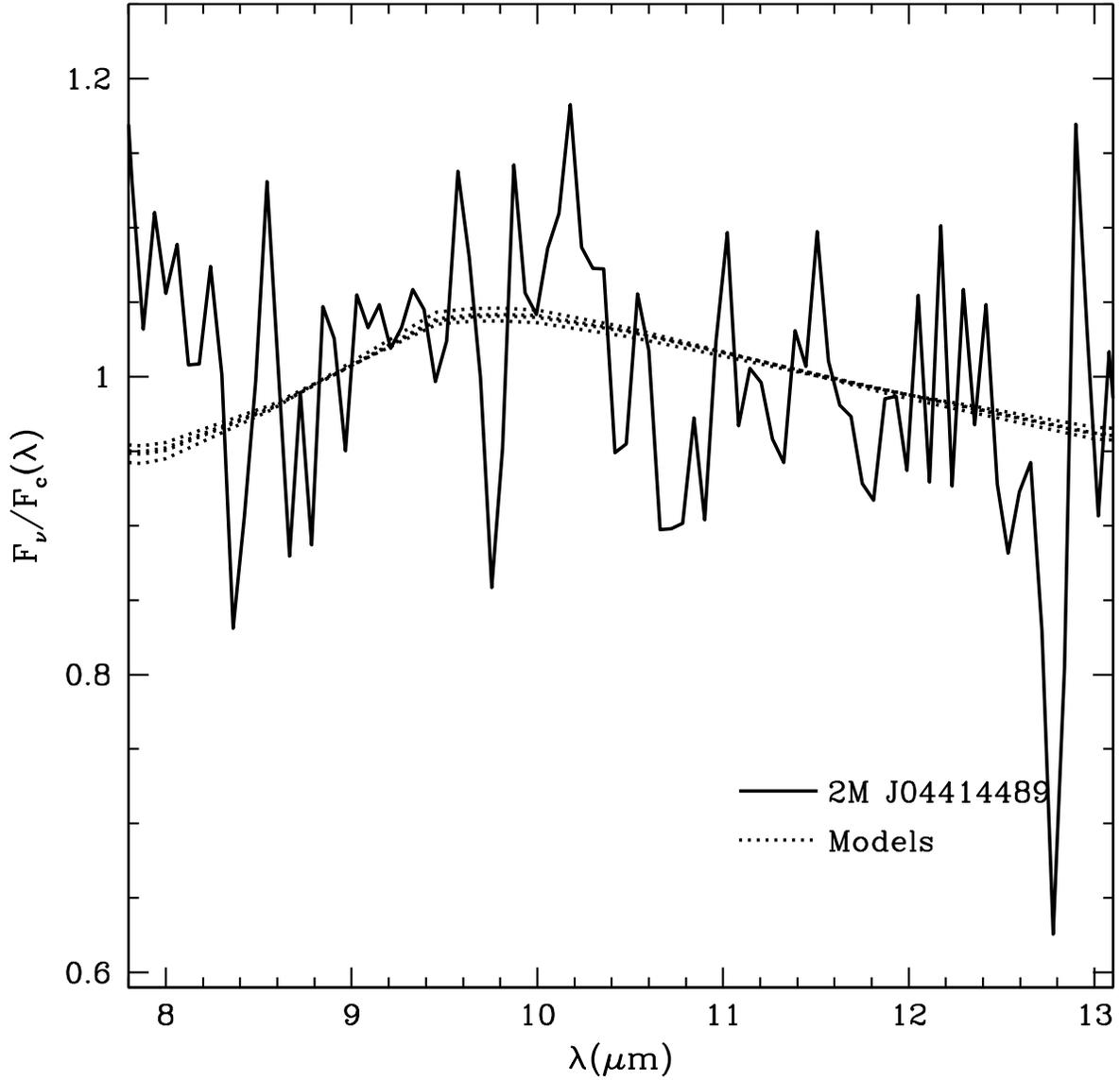}
\caption{Spectrum near 10~\micron\ for 2M~J04414489 normalized to the
continuum surrounding the wavelengths of the silicate feature. For
comparison, we show the normalized silicate emission produced by the
models in Table~\ref{tbl-1}. The order of the models from strongest to
weakest emission peak is B, A, C, and D.}
\label{fig:fig2}
\end{figure}

We compare the silicate emission predicted by our best-fit models to the
IRS data for 2M~J04414489 in Figure~\ref{fig:fig2}. We normalize the IRS
spectrum and the models with a continuum interpolated from a third-order
polynomial constructed by fluxes between $\sim5.2$ and
$\sim7.9$~\micron\ and between $\sim13$ and 16~\micron\
\citep[e.g.][]{Furlan06}. The strength of the silicate feature relative
to the continuum is obtained by subtracting the continuum to the
feature, integrating this residual flux and normalize it to the
integrated continuum flux between $8$ and $13$~\micron\
\citep{Furlan06}. The strength of the observed feature is 1.03, whereas
the strengths of our models features range from 1.024 to 1.032. Thus,
the observed and modeled silicate features are similar, but also, they
are extremely weak. The absence of detectable silicate emission in the
data indicates that grains in the disk atmosphere have grown to sizes of
$5\lesssim a_{max}<1000$~\micron, which are larger than the
typical ISM grains.

The dust masses of our final set of disk models are listed in
Table~\ref{tbl-1}, using $\zeta_{std}=0.0065$ \citep{DL84}. The masses
of the models range from $\sim3$ to $\sim5$~$M_{Moon}$, which are more
than an order of magnitude lower than the values estimated from
millimeter measurements of other brown dwarf disks in Taurus
\citep{Scholz06}. This result is not surprising given that close binary
stars exhibit low disk masses compared to single stars and wide binaries
\citep{Jensen94}.

\subsection{Size of the Mid-IR Emission Region}

The mid-IR emission from an isolated disk irradiated by a M8.5 brown
dwarf typically forms within $R_{MIR}\lesssim5$~AU. The exact size of
this region depends the size of the grains \citep{KS07}, but also on the
dust mass surface density through the variation of $\epsilon$, $\dot{M}$
or $\alpha$. We find that decreasing $\alpha$ or increasing $\dot{M}$ or
$\epsilon$ result in a wider emitting region since those variations
increase the irradiation surface and the impinging irradiation flux at
any given radius. On the other hand, decreasing $a_{max}$ will increase
the optical depth to the incident radiation at the upper layers,
which in turn extends vertically outward the irradiation surface,
allowing more distant annuli to contribute to the emission.

\begin{figure}
\epsscale{1}
\plotone{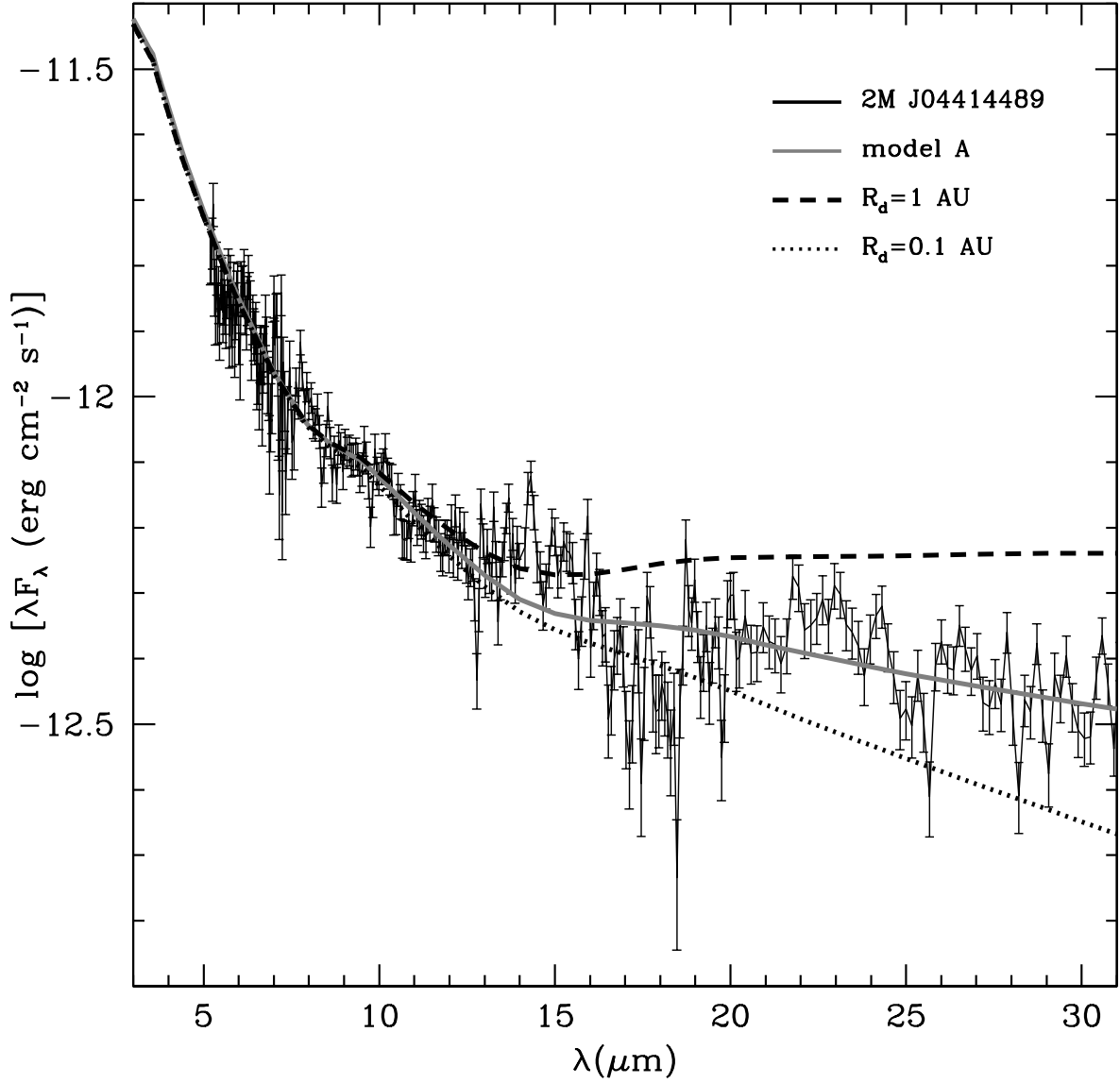}
\caption{SED of the young brown dwarf 2M~J04414489 compared to model A
($R_d=0.2$~AU) and two models that have the same parameters as model I
except for different values of $R_d$.}
\label{fig:fig3} 
\end{figure}

If the outer radius of the disk is smaller than the nominal radius of the
mid-IR emitting region, then the mid-IR SED is sensitive to the value of
the outer radius. For instance, the radii for all of the models in
Table~\ref{tbl-1} are constrained to be $0.2\lesssim R_d\lesssim0.3$~AU
by the mid-IR SED of 2M~J04414489, which is smaller than the expected
radius given by the binary tidal interaction ($R_{truncation}\sim3$--7
AU). However, if the eccentricity of the binary is high enough, the
periastron would be close to 1~AU if a semi-major axis of 15~AU is still
assumed. The dependence of the predicted SED on the outer radius is
illustrated in Figure~\ref{fig:fig3} for model A. Although the disk
radii for our models of 2M~J04414489 are roughly similar to that derived
for the circumprimary disk of SR 20 \citep[0.39~AU, for a G7 primary and
a binary projected separation of $5.3-9.9$~AU,][]{McClure08}, the former
has a redder SED (i.e., less affected by truncation) because the mid-IR
emitting region of a brown dwarf is much smaller than that of a star.

\section{Discussion}

We find that the dust in the disk around 2M~J04414489 is relatively
evolved for the age of the system ($\sim1$~Myr). Based on the slope of
its SED and the absence of silicate emission, most of the initial
atmospheric grains have settled towards the midplane
($0.01\le\epsilon\le0.1$) and the remaining grains may have experienced
significant growth ($5\lesssim a_{max}<1000$~\micron),
representing one of $\sim2$ brown dwarf disks in Taurus that show
evidence of accelerated dust evolution \citep[see KPNO 6\footnote{For
the other objects reported by \citet{Riaz09}, silicate emission is
detected at weak levels in the same spectra that were re-reduced (Furlan
et al. 2011, in preparation).},][]{Riaz09}. This degree of dust
evolution has been observed primarily for brown dwarfs at ages of
5--10~Myr \citep{Scholz07,Morrow08}.

Some mechanisms to explain grain growth and settling in disks around
either single or multiple late-type M-dwarfs have been put forward.
\citet{M03} suggested that binary interaction may accelerate the dust
evolution through an enhanced vertical stirring of grains from the
midplane. However, \citet{Pascucci08} and Furlan et al. (2011, in
preparation) found no significant differences between the silicate
feature for early type single and binary star systems in Taurus,
revealing that binary interaction may not affect the inner disk region,
although the outermost region is indeed influenced. The interaction
truncates the disk to a given radius \citep{PP77,AL94}, decreasing the
emission from a certain wavelength range. For instance, the
submillimeter and millimeter emission from the circumprimary disk is
diminished for projected binary separations of $\sim50$---100~AU
\citep{Jensen94,Jensen96}, while for smaller separations
($\lesssim50$~AU), the mid-IR emission of the outer disk decreases with
respect to that of the innermost region \citep{McClure10}. The mid-IR
emission of J04414489 is the result of this truncation, whereas the
fluxes near $\sim10$~\micron\ are unlikely influenced by the binary
interaction. Instead, the inner region may be probably affected by the
turbulence as most of the eddie dissipation occurs within those annuli
($\lesssim0.1$~AU), which represents $\sim30$---50\% of the total
radial extension of the disk.

Grain growth and settling should be efficient in regions
close to the star, where temperatures and densities are larger, and the
turbulent movements are higher \citep[][references
therein]{BHN00,DD04,NN06}. For this weakly turbulent disk
($\alpha=10^{-5}$), the coagulation rate may be slowed down, but also
the relative velocities of the grains, thus preventing their
fragmentation. The smaller velocity gradients can work to preserve the
large grains once formed. A radial mixing mechanism may exist that moves
evolved grains outward and replenishes the inner disk with primordial
dust \citep[e.g. turbulent diffusion, convection, gravitational torques,
among others, see][]{G01,BM02,B04}. The primordial dust reservoir is
limited by the outer radius of a disk. If a disk is outwardly truncated
to a small enough radius (e.g., $R_d<R_{MIR}$), the outermost region
will contain evolved grains, which return to the innter region to be
processed further. The characteristic timescale for this radial mixing
is approximately the viscous timescale \citep{BM02}, which for an
$\alpha$-disk, $t_{RM}=R_0^{2}\Omega_K(R_0)/3\alpha c^2_s(T_c)$. Here,
$c_s(T_c)$ is the sound speed at the midplane and $\Omega_K$ the
Keplerian angular speed, both evaluated at $R_0$ (the initial outer
radius of the disk). We evaluate $t_{RM}$ at the outer radii listed in
Table~\ref{tbl-1} assuming a central temperature at the midplane of
$\sim60$---80~K. The mixing timescale derived is $\sim1$---2 Myr, which
is comparable to the age of 2M J04414489. The radial mixing may be at an
earliest stage, but its influence over the dust processing cannot be
neglected. Therefore, low levels of turbulence combined with a highly
truncated disk could explain the deficit of primordial dust that we
observe in the disk of 2M J04414489.

\section{Conclusions}
We have presented a mid-IR spectrum of the binary brown dwarf
2M~J04414489 in the Taurus star-forming region. The spectrum exhibits
excess emission that we attribute to a circumprimary disk, but silicate
emission at 10~\micron\ is not present, indicating that significant
grain growth has occurred, leading to a disk on which small grains no
longer exist ($5\lesssim a_{max}<1000$~\micron). This is one of
$\sim2$ brown dwarf disks at such a young age ($\sim1$~Myr) that have
been found to lack silicate emission. Our models of the SED of
2M~J04414489 suggest that the outer radius of its disk is rather small
($R_d\sim0.2$--0.3~AU), which may reflect truncation by the binary
companion. The absence of an outer disk with a reservoir of primordial
grains, combined with weak turbulence, could explain the advanced grain
growth in the inner disk of 2M~J04414489. 

\acknowledgements 
We would like to thank the anonymous referee for his/her valuable
comments. We acknowledge support from grant IN112009 from PAPIIT-DGAPA
UNAM (P.D. and L.A.), grant NNX08AH94G from NASA (N.C. and L.A.), and
grant AST-0544588 from the National Science Foundation (K. L.). Part of
the numerical calculations were performed on the cluster at CRyA-UNAM,
acquired through CONACyT grant 36571-E to Enrique V\'azquez-Semadeni.
The Center for Exoplanets and Habitable Worlds is supported by the
Pennsylvania State University, the Eberly College of Science, and the
Pennsylvania Space Grant Consortium.

\begin{deluxetable}{lrrccccr} 
\tablecolumns{8} 
\tablewidth{0pt}
\tablecaption{Models\tablenotemark{a}\label{tbl-1}}
\tablehead{
\colhead{Model\tablenotemark{b}} & \colhead{$\epsilon$} & \colhead{$a_{max}$} &
\colhead{z$_{wall}$} & \colhead{$R_d$} & \colhead{$M_{dust}$} &
\colhead{$\chi^2_{r}$} & \colhead{$t_{RM}$} \\
\colhead{} & \colhead{} & \colhead{(\micron)} & \colhead{(H$_p$)} &\colhead{(AU)} &
\colhead{(M$_{Moon}$)} & \colhead{} & \colhead{($10^{6}$~yrs)}
}
\startdata 
A & 0.1 & 10 & 3.7 & 0.2 & 3.2 & 3.7 & 1.2\\
B & 0.1 & 5 & 3.6 & 0.2 & 3.2 & 3.8 & 1.2\\
C & 0.1 & 100 & 3.9 & 0.3& 5.2 & 3.9 & 1.8\\
D & 0.01 & 5 & 3.8 & 0.3 & 5.7 & 3.9 & 1.9 \\
\enddata 
\tablenotetext{a}{The listed parameters are: the dust-depletion
paramenter $\epsilon$, the maximum grain radius of the dust population
in the disk atmosphere $a_{max}$, the height of the wall $z_{wall}$,
the outer disk radius $R_d$, the total dust content of the disk
$M_{dust}$, the reduced-$\chi^2$ value of the model, and the
characteristic timescale for the radial mixing, evaluated at $R_d$,
$t_{RM}$.}
\tablenotetext{b}{$\alpha=10^{-5}$,
$\dot{M}=10^{-11}$~M$_{\sun}$~yr$^{-1}$, $i=20\arcdeg$,
$R_{in}=6$~R$_{*}$ for all models.}
\end{deluxetable}

\end{document}